A Scoring System for Continuous Glucose Monitor Data

Edward F. Aboufadel, PhD.

Department of Mathematics

Grand Valley State University

May 2013 version

Contact information:

Prof. Edward Aboufadel

Dept. of Mathematics

Grand Valley State University

Allendale, MI  49401

Phone:  616-331-2445

Fax:  616-331-3120

e-mail:  aboufade@gvsu.edu

**Abstract:** As continuous glucose monitors (CGMs) are used increasingly by diabetic patients, new and intuitive tools are needed to help patients and their physicians use these streams of data to improve blood glucose management. In this paper, we propose a daily CGM Score which can be calculated from CGM data. The calculation involves assigning grades and scores to 80-minute periods of CGM data, and then aggregating the results. Scores for an individual patient, or among a set of patients, can then be compared and contrasted, and longitudinal studies of CGM data can also be accomplished.

As continuous glucose monitors (CGMs) are used increasingly by diabetic patients, how can those patients and their physicians apply these streams of data to improve blood glucose management? Currently, a type-1 diabetic patient receives a low-frequency diagnostic every three months: the $A_{1c}$ test. With a CGM, data can be recorded at a high frequency of every five minutes. In this paper, we propose a "CGM Score" which can be used at intermediate frequencies – daily, weekly, etc. It is an intuitive tool that can help patients and physicians understand long-term trends in glucose management.

In a 24-hour period, a properly performing CGM system will record 288 glucose readings – one every five minutes – and a variety of dynamics may be observed, as indicated in Figure 1 to 4. With CGM readings, a diabetic patient can use the most current reading and readings from the past hour or so to make decisions about extra insulin or carbohydrates. This is the basis of the PID (position-integral-derivative) algorithm **[1]** for an artificial pancreas. What we are considering in this paper is something different: we define a diagnostic tool to provide a *post hoc* analysis of CGM data which is much richer than the $A_{1c}$ number. The CGM Score captures volatility in the CGM time series, and for certain patients, can reveal issues with glucose management that are masked by a good $A_{1c}$ number. We will show how to calculate a CGM Score for any 24-hour period. Scores for an individual patient, or among a set of patients, can then be compared and contrasted.

We developed our proposed CGM Score using a publicly-available database[1] of CGM data of 451 patients that was collected as part of a recent study by the Jaeb Center for Health Research **[3]**. The study involved type-1 diabetes patients wearing a CGM for 6 or 12 months. In an initial analysis of the database, we identified 42,799 days which had at least 209 out of 288 possible CGM readings, coming from 346 patients, and these are the days and patients used in this paper. For the days that were used, when CGM readings were missing, the previous known reading was used as a reasonable proxy.

To calculate a CGM Score for a 24-hour period, we begin by dividing the 288 readings into 18 periods of 16 readings, or 80 minutes, each. In the signal processing literature, it is common to use signals whose length is a power of 2 (e.g., when wavelet filters are applied **[2]**). We will evaluate each 80 minute period separately, and then combine these 18 evaluations into a total score for the day. Within an 80-minute period, a type-1 diabetic patient may see a spike up or down, but it is unlikely that both will occur in that short of a time period.  Also, a set of 18 evaluations for a 24-hour period appears to be sufficient to capture the dynamics of the day, based on our efforts to develop this score.

---

[1] http://diabetes.jaeb.org/RT_CGMRCTProtocol.aspx

We assess each 80-minute period by taking into account (1) the glucose level in that period, (2) whether the glucose numbers are increasing or decreasing, and (3) the rate of increase or decrease. For (1), we take the average of the first five (out of 16) glucose entries for the period, as (2) and (3) will capture a sense of the rest of the entries.  For (2) and (3), we simply calculate the difference between the last entry in the 80-minute period and the first, which we will call Δ.  Based on these calculations, we assign a letter grade to the 80-minute period, with letters closer to "A" corresponding to "better" glucose management.  Here is how we assign these grades:

|  | flat | trend up | trend down | spike up | spike down |
| --- | --- | --- | --- | --- | --- |
| Average of first five entries | -20 ≤ Δ ≤20 | 21 ≤ Δ ≤99 | -99 ≤ Δ ≤21 | 100 ≤ Δ | Δ ≤-100 |
| more than 240 | L | M | G | W | H |
| 140-239 | F | K | E | R | P |
| 70-139 | A | B | C | Q | Y |
| less than 70 | V | D | X | J | Z |

The next step is to convert the 18-letter code into a single score for the day.  For each grade, we assign a number that seems to reasonably capture which dynamics are "better" or "worse", with smaller numbers corresponding to higher letter grades.  In general, we instantiate preferences that trending is better than spiking, and that highs are better than lows while being in-range is better than either.  Fixing a low gets a good grade, and we want to give a reasonable score to a high if it is being corrected.  With all these ideas in the mind, we have the following scores:

|  | flat | trend up | trend down | spike up | spike down |
| --- | --- | --- | --- | --- | --- |
| Average of first five entries | -20 ≤ Δ ≤20 | 21 ≤ Δ ≤99 | -99 ≤ Δ ≤21 | 100 ≤ Δ | Δ ≤-100 |
| more than 240 | 50 | 60 | 40 | 80 | 45 |
| 140-239 | 25 | 45 | 20 | 65 | 60 |
| 70-139 | 0 | 2 | 2 | 60 | 95 |
| less than 70 | 75 | 10 | 95 | 45 | 100 |

Examples of codes and scores for individual days can be found in Figures 1 to 4.  (Note:  In the Jaeb database, the day of recording of the CGM data is given using the Excel format, where January 1, 1900 at 12 midnight is "1".)  Using our proposed CGM Score on the Jaeb database, we make the following observations.

First, it is important to note that the CGM score is an ordinal measure, meaning that it can indicate if one CGM is better than another, but it makes no sense to say, for instance, that a score of 200 is twice as good as a score of 400.  For that reason, computing means of CGM scores does not make a lot of sense.  However, percentiles and medians are valid calculations to make to understand the data better.

For the days in the database, the CGM Scores ranged from 0 to 1179, and the median score was 387.  A histogram of the distribution of scores is in Figure 5, and the cut-off scores for each percentile can be found in Table 1.

Figure 6 has a graph of the median CGM Score for each patient compared to an $A_{1C}$ number reported in the Jaeb database for each patient, and we see a good correlation, as we should.   However, what is interesting is that there are a number of patients who have a relatively low $A_{1C}$ number but a relatively high CGM Score.  In some cases, it is because the amount of CGM data reported is small.  There are other cases, however, such as patient #43 in the Jaeb database, with interesting dynamics.  This patient has a reported $A_{1C}$ of 6.2%, but her median CGM Score is 412, which is the 40$^{th}$ percentile.  The CGM graphs for this patient (two examples can be found in Figure 7) suggest some explanations.  One is that while the average blood glucose may be low, there is a lot of volatility, with spikes high and low, which yields a larger CGM Score.  Another possible explanation is that this patient might be maintaining low blood sugar that is potentially dangerous, which keeps $A_{1C}$ low, but a higher CGM Score due to the "V" grades.  Comparisons of $A_{1C}$ numbers with CGM Scores may lead to good conversations between physicians and patients about glucose management practices.

Physicians and patients can also investigate how the CGM Score varies over time.  For example, in Figure 8 is a histogram of CGM Scores for patient #14 in the database, and this patient appears to be managing her blood glucose well, with most scores under 200 and an obvious drop-off as the daily score increases from 200 to past 500.  On the other hand, the histogram for patient #193 (see Figure 9), reflects poor glucose control, and this patient had a reported $A_{1C}$ of 8.9%.  Another way to study CGM Scores over time would be to create a graph of the scores from day to day, and then compute a regression line.  A line that is sloping downwards would indicate a trend that CGM Scores are decreasing, which would be a desirable result.  Figure 10 has an example with patient #63.

There are several other papers on measuring or evaluating blood glucose variability.  Perhaps the best known is the MAGE measure (Mean Amplitude of Glycemic Excursions), which compares the standard deviations of CGM readings with the differences between consecutive readings **[4]**.  The ADRR (Average Daily Risk Range) variability index provides a number that varies gradually over time which captures how quickly blood glucose numbers have been changing from high to low, or low to high **[5]**.  A stochastic model has also been developed **[6]**, and two recent papers with an overview on the subject are **[7]** and **[8]**.  We believe the system proposed in this paper, which involves computing grades and scores of 80-minute periods of CGM data, will be easier to explain to patients.

Could CGM Scores actually be used by health-care professionals in the future?  In order to do so, a clinical study would be required, collecting CGM data to be analyzed by a panel of experts to refine the scoring system.  An expert-approved system could then be applied to compare scores with demographic data such as age, time from diagnosis, education level of the primary caregiver, and gender.  Recommended advice based on CGM Scores would need to be settled upon by health-care researchers and practicing physicians.  We hope this paper will be a catalyst for such a study and for further development of methods of analysis for CGM data for diabetic patients.

**References**

**[1]**  Panteleon AE, Loutseiko M, Steil GM, Rebrin K: Evaluation of the effect of gain on the meal response of an automated closed-loop insulin delivery system.  *Diabetes* 55: 1995-2000, 2006.  doi: 10.2337/db05-1346

**[2]**  Aboufadel EA, Schlicker SJ:  Wavelets (Introduction).  In *Encyclopedia of Physical Science and Technology*, 3rd edition. Meyers R, Ed.  San Diego, Academic Press, 2001, p.773-788.  doi: 10.1016/B0-12-227410-5/00823-1

**[3]**  JDRF CGM Study Group:  JDRF randomized clinical trial to assess the efficacy of real-time continuous glucose monitoring in the management of type 1 diabetes: research design and methods.  *Diabetes Technology and Therapeutics* 10(4): 310-321, 2008.  doi: 10.1089/dia.2007.0302

**[4]**   Service, FJ, Molnar GD, Rosevear JW, Ackerman  E, Gatewood LC, Taylor WF:  Mean amplitude of glycemic excursions, a measure of diabetic instability. *Diabetes*, 19:644-655, 1970.

**[5]**   Kovatchev BP, Otto E, Cox D, Gonder-Frederick L, Clarke W:  Evaluation of a new measure of blood glucose variability in diabetes. *Diabetes Care* 29(11):2433-2438, 2006.  doi: 10.2337/dc06-1085

**[6]**  Magni P, Bellazzi R:  A stochastic model to assess the variability of blood glucose time series in diabetic patients self-monitoring.  *IEEE Transactions on Biomedical Engineering* 53(6):977-985, 2006.  doi: 10.1109/TBME.2006.873388

**[7]**  Rahaghi FN, Gough DA:  Blood glucose dynamics.  *Diabetes Technology & Therapeutics* 10(2):81-94, 2008. doi:10.1089/dia.2007.0256

**[8]**  Sparacino G, Facchinetti A, Cobelli C.  "Smart" continuous glucose monitoring sensors: on-line signal processing issues. *Sensors* 10(7):6751-6772, 2010. doi:10.3390/s100706751

Biographical Information:  Edward Aboufadel received his Ph.D. in Mathematics from Rutgers University in 1992. He is a Professor of Mathematics at Grand Valley State University, Allendale, Michigan.

**Table 1**

| **99** | **98** | **97** | **96** | **95** | **94** | **93** | **92** | **91** | **90** |
|---|---|---|---|---|---|---|---|---|---|
| 33 | 59 | 77 | 91 | 102 | 114 | 124 | 133 | 143 | 151 |
| **89** | **88** | **87** | **86** | **85** | **84** | **83** | **82** | **81** | **80** |
| 159 | 167 | 173 | 180 | 187 | 192 | 199 | 205 | 211 | 217 |
| **79** | **78** | **77** | **76** | **75** | **74** | **73** | **72** | **71** | **70** |
| 223 | 229 | 234 | 239 | 245 | 250 | 255 | 260 | 264 | 269 |
| **69** | **68** | **67** | **66** | **65** | **64** | **63** | **62** | **61** | **60** |
| 274 | 279 | 283 | 288 | 293 | 298 | 303 | 307 | 312 | 317 |
| **59** | **58** | **57** | **56** | **55** | **54** | **53** | **52** | **51** | **50** |
| 321 | 326 | 331 | 336 | 340 | 345 | 350 | 354 | 359 | 364 |
| **49** | **48** | **47** | **46** | **45** | **44** | **43** | **42** | **41** | **40** |
| 369 | 374 | 378 | 384 | 388 | 393 | 399 | 404 | 409 | 414 |
| **39** | **38** | **37** | **36** | **35** | **34** | **33** | **32** | **31** | **30** |
| 419 | 424 | 429 | 435 | 440 | 446 | 451 | 457 | 462 | 468 |
| **29** | **28** | **27** | **26** | **25** | **24** | **23** | **22** | **21** | **20** |
| 474 | 479 | 485 | 491 | 497 | 503 | 510 | 517 | 524 | 531 |
| **19** | **18** | **17** | **16** | **15** | **14** | **13** | **12** | **11** | **10** |
| 539 | 546 | 554 | 561 | 569 | 577 | 587 | 596 | 606 | 617 |
| **9** | **8** | **7** | **6** | **5** | **4** | **3** | **2** | **1** | |
| 630 | 642 | 657 | 672 | 690 | 709 | 734 | 767 | 817 | |

Percentile cut-off scores for CGM Scores

**Figure 1**

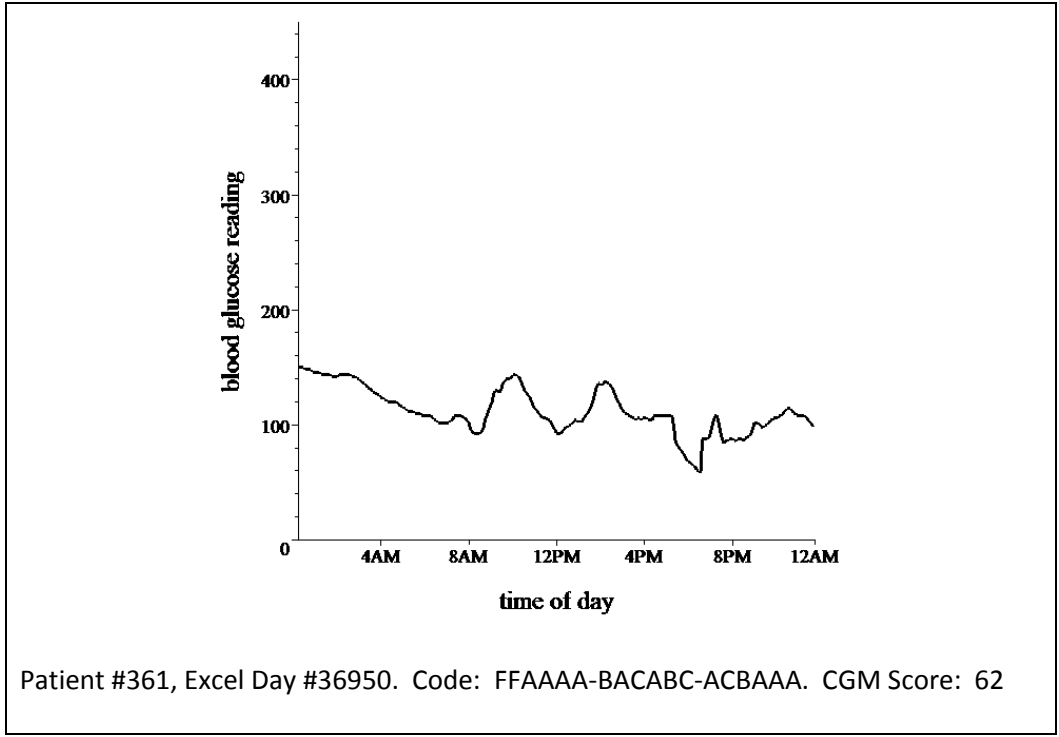

Patient #361, Excel Day #36950.  Code:  FFAAAA-BACABC-ACBAAA.  CGM Score:  62

**Figure 2**

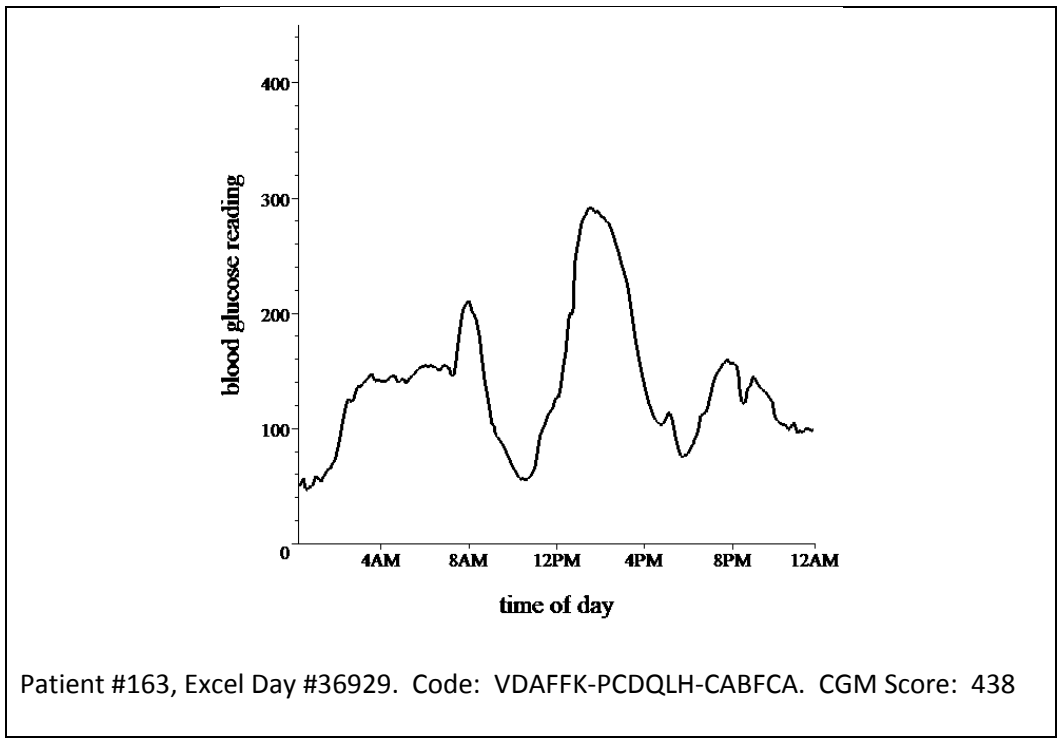

Patient #163, Excel Day #36929.  Code:  VDAFFK-PCDQLH-CABFCA.  CGM Score:  438

**Figure 3**

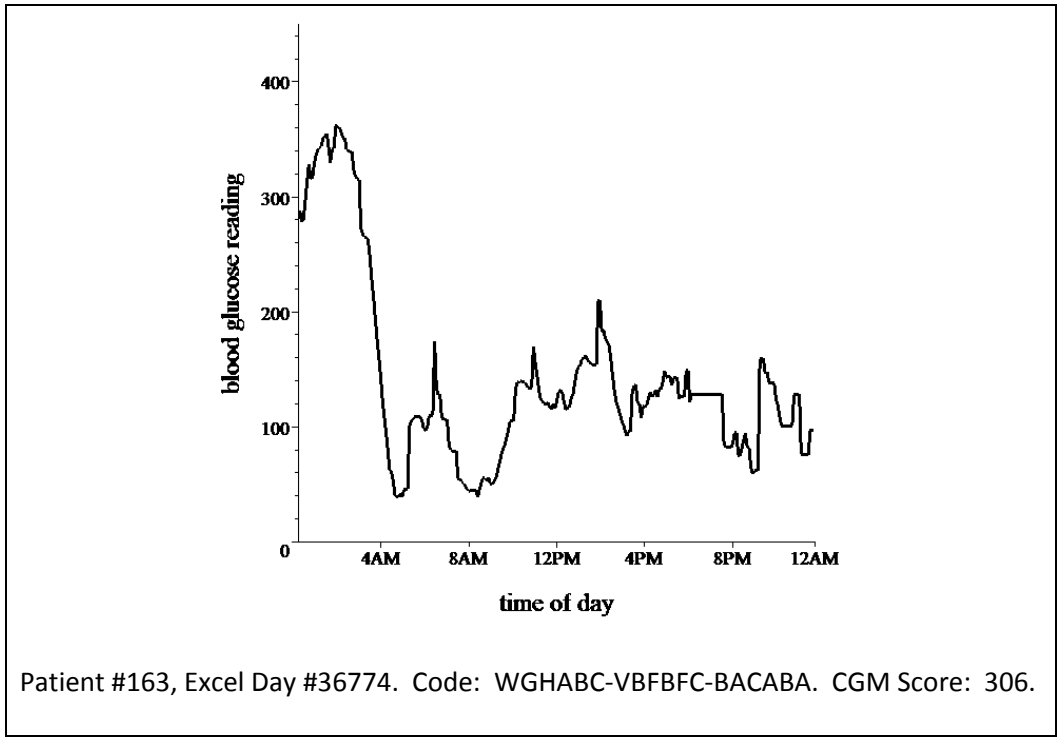

Patient #163, Excel Day #36774.  Code:  WGHABC-VBFBFC-BACABA.  CGM Score:  306.

**Figure 4**

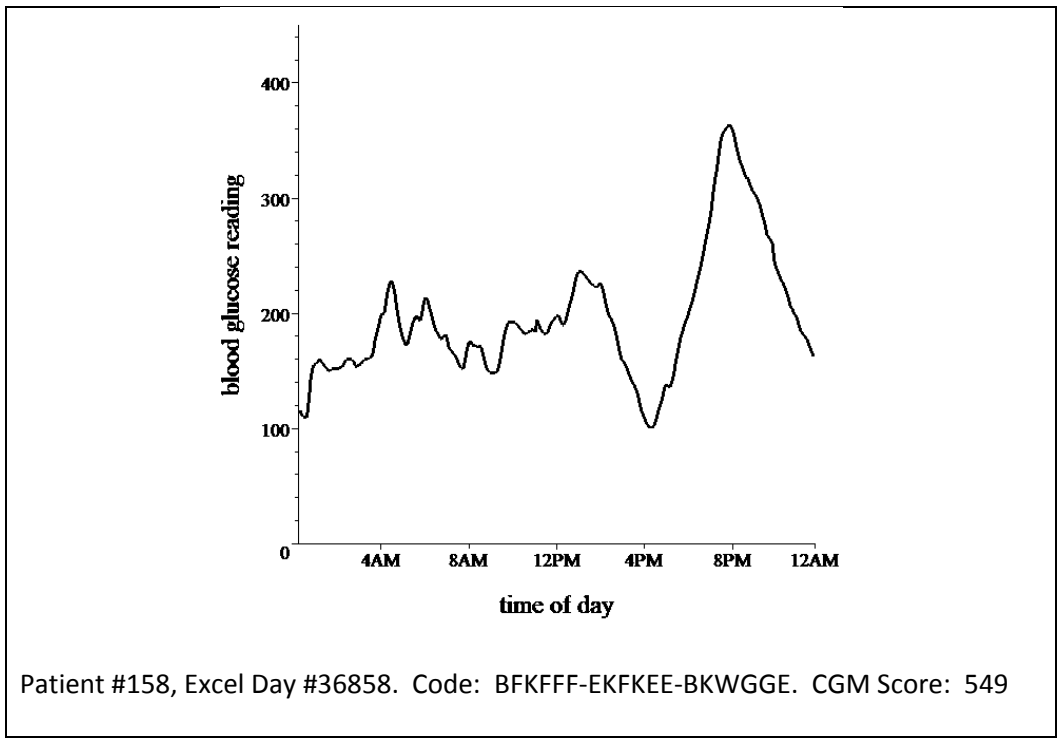

Patient #158, Excel Day #36858.  Code:  BFKFFF-EKFKEE-BKWGGE.  CGM Score:  549

**Figure 5**

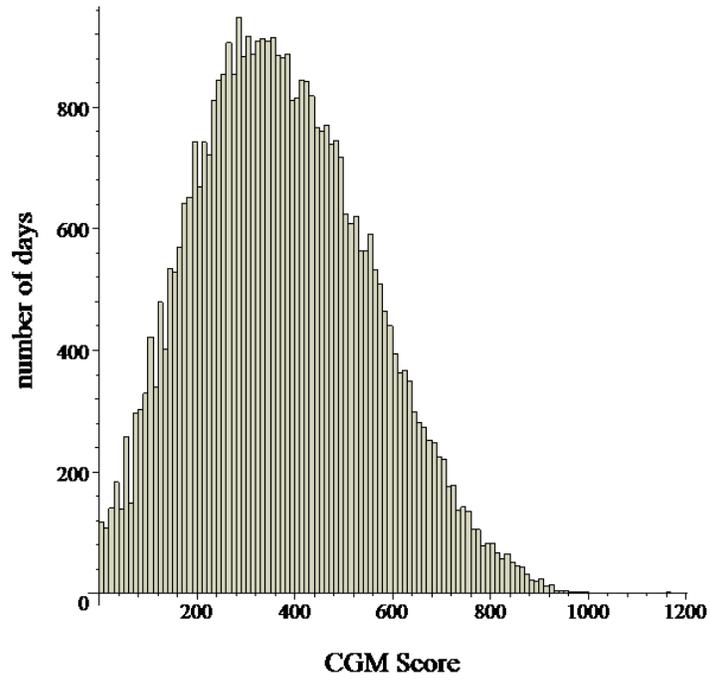

Distribution of CGM Scores in the Jaeb Database

**Figure 6**

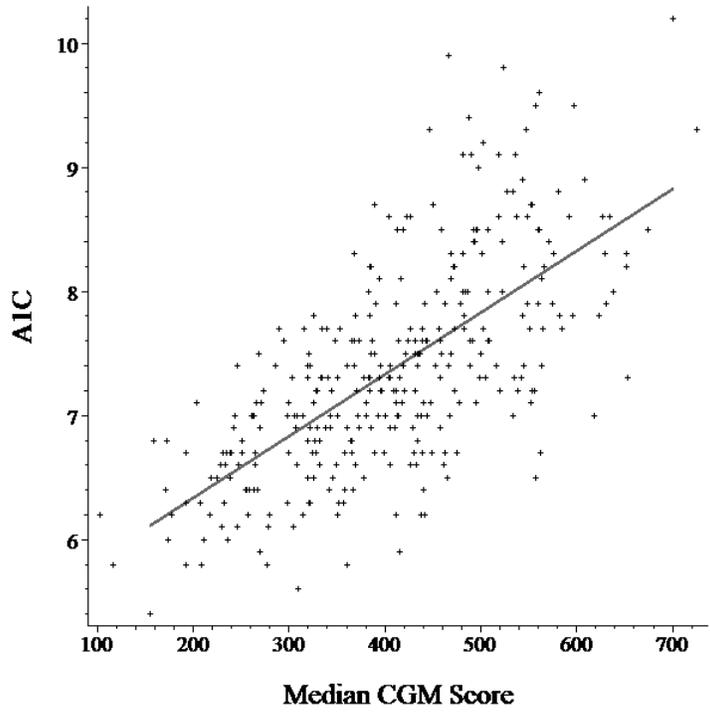

Comparison of Median CGM Scores with $A_1C$

**Figure 7:**

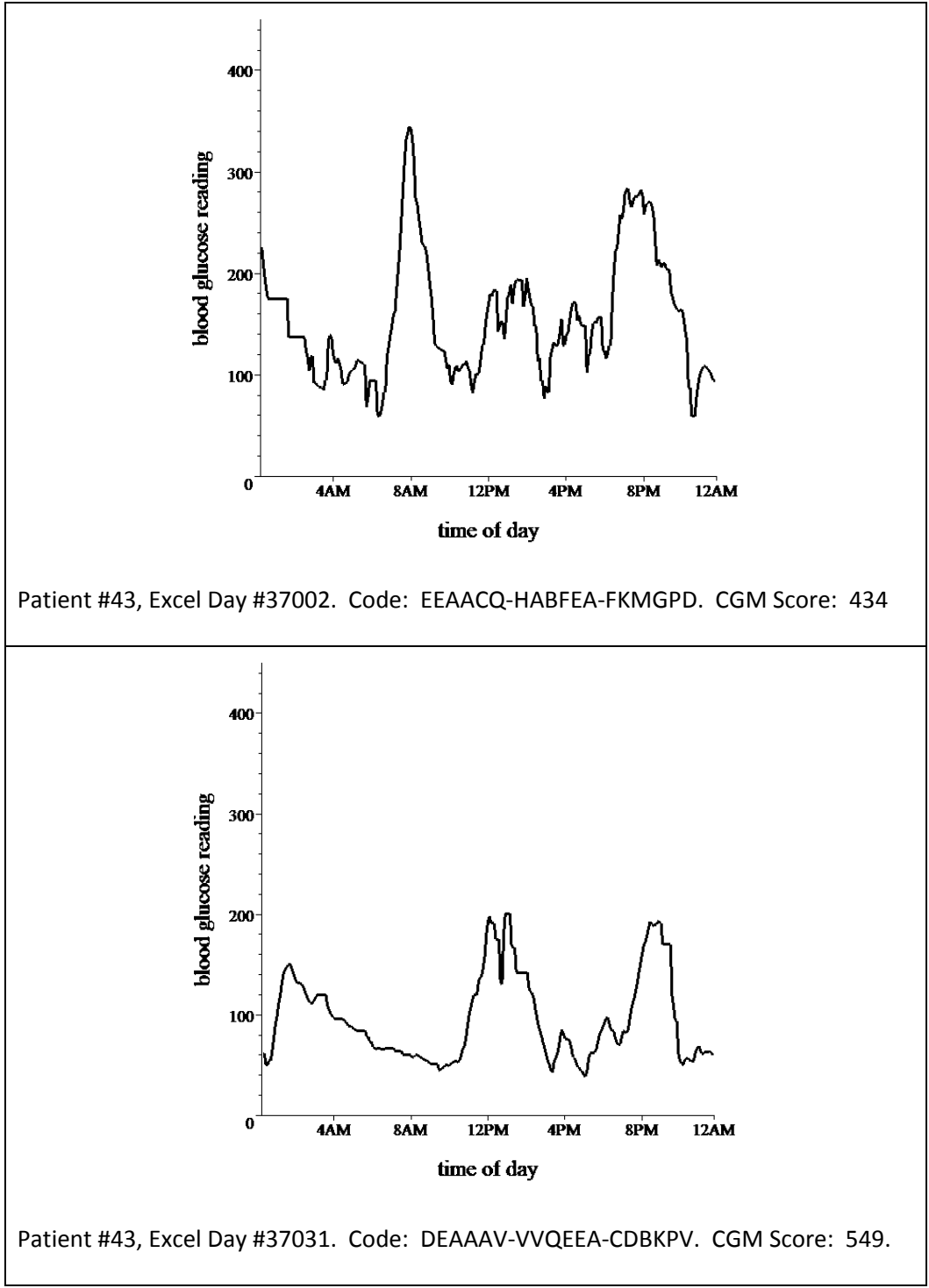

Patient #43, Excel Day #37002. Code: EEAACQ-HABFEA-FKMGPD. CGM Score: 434

Patient #43, Excel Day #37031. Code: DEAAAV-VVQEEA-CDBKPV. CGM Score: 549.

**Figure 8:**

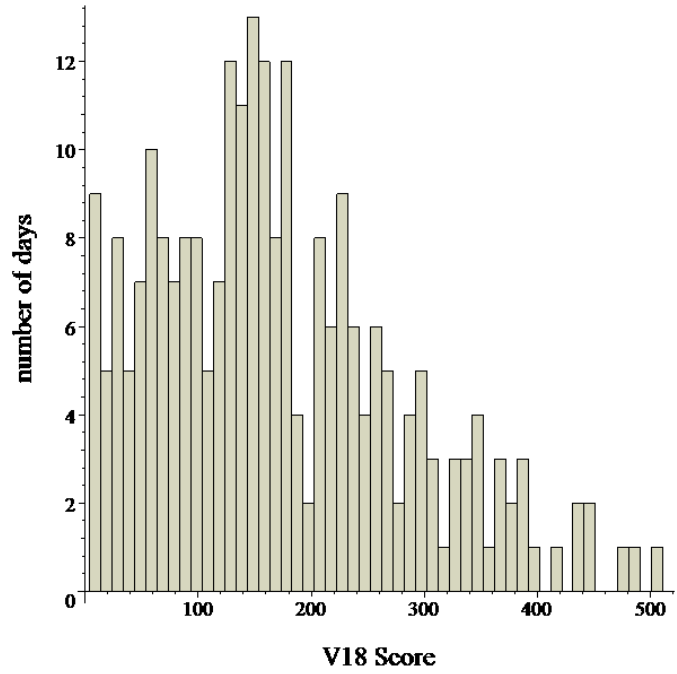

Histogram of CGM Scores for Patient #14

**Figure 9:**

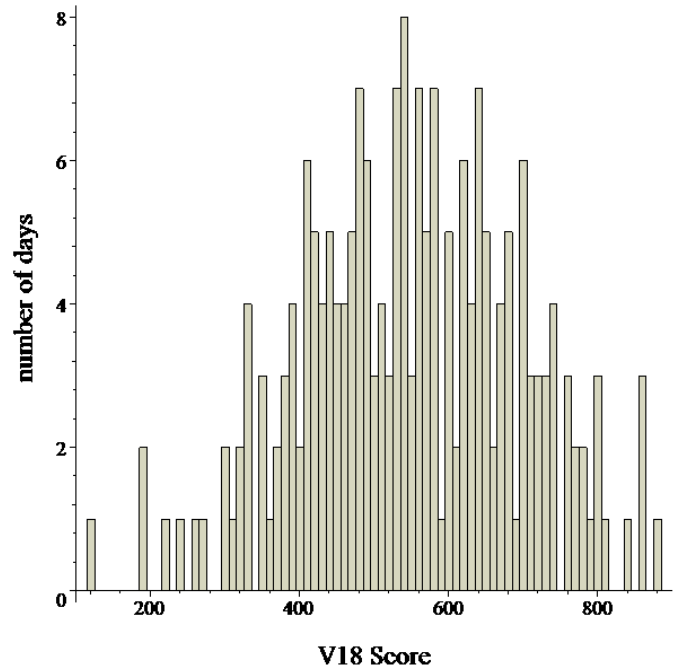

Histogram of CGM Scores for Patient #193

**Figure 10**

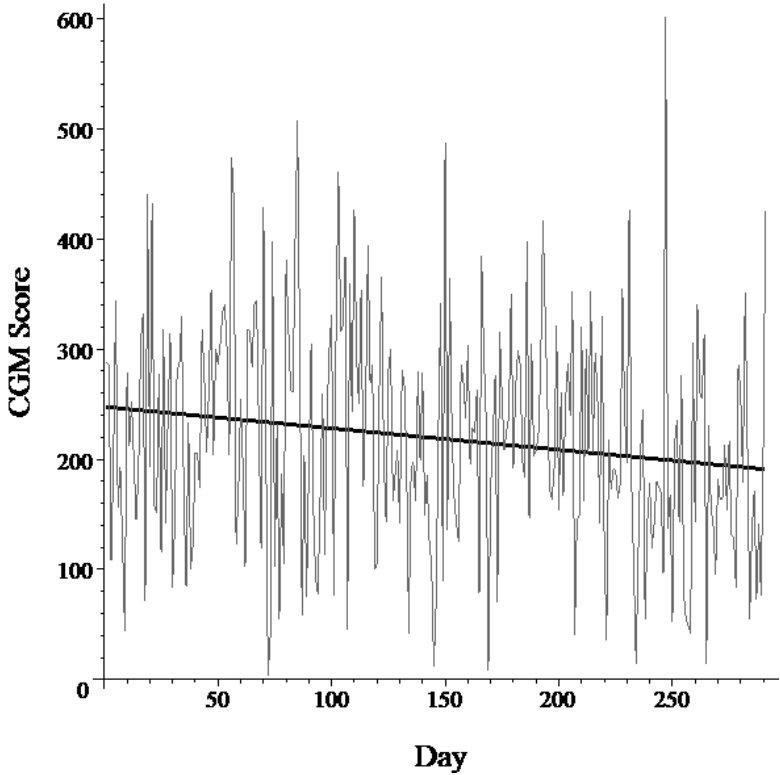

Day-to-day CGM Scores for Patient #63